\documentclass[12pt]{article}
\usepackage{amssymb,amsmath}
\usepackage{graphicx}
\usepackage{subfig}
\usepackage{color}
\usepackage{slashed}
\usepackage{enumerate}
\usepackage{dsfont}

\newcommand{\beq}{\begin{equation}}
\newcommand{\eeq}{\end{equation}}
\newcommand{\bea}{\begin{eqnarray}}
\newcommand{\eea}{\end{eqnarray}}
\newcommand{\bec}{\begin{center}}
\newcommand{\eec}{\end{center}}
\newcommand{\bei}{\begin{itemize}}
\newcommand{\eei}{\end{itemize}}

\def\10{$SO(10)$}
\def\21{SU(2) $\otimes$ U(1) }

\def\422{$SU(4) \otimes SU(2) \otimes SU(2)$}
\def\321{SU(3) $\otimes$ SU(2) $\otimes$ U(1)}

\def\lsim{\raise0.3ex\hbox{$\;<$\kern-0.75em\raise-1.1ex\hbox{$\sim\;$}}}
\def\gsim{\raise0.3ex\hbox{$\;>$\kern-0.75em\raise-1.1ex\hbox{$\sim\;$}}}

\begin{document}

\begin{titlepage}
\begin{flushright}
  DO-TH 11/30
\end{flushright}
  \newcommand{\AddrDOR}{{\sl \small Facult\"at f\"ur Physik, Technische 
      Universit\"at Dortmund\\ D-44221 Dortmund, Germany}}
  \newcommand{\AddrSISSA}{{\sl \small SISSA and INFN, Sezione
      di Trieste,\\ Via Bonomea 265, 34136 Trieste, Italy}}
  \newcommand{\AddrLiege}{{\sl \small IFPA, Dep. AGO, 
      Universite de Liege, Bat B5,\\ \small \sl Sart
      Tilman B-4000 Liege 1, Belgium}}
\vspace*{1.5cm}
\begin{center}
   \textbf{\large Leptogenesis in flavor models}\\[3mm]
   \textbf{\large with type I and II seesaws}
  \\[10mm]
  D. Aristizabal Sierra$^{a,}$\footnote{e-mail address: daristizabal@ulg.ac.be},
  F. Bazzocchi$^{b,}$\footnote{e-mail address: fbazzo@sissa.it},
  I. de Medeiros Varzielas$^{c,}$\footnote{e-mail address: ivo.de@udo.edu}
  \vspace{0.8cm}
  \\
  $^a$\AddrLiege.\vspace{0.4cm}\\
  $^b$\AddrSISSA.\vspace{0.4cm}\\
  $^c$\AddrDOR.\vspace{0.4cm}  \\
\end{center}
\vspace*{0.5cm}
\begin{abstract}
  In type I seesaw models with flavor symmetries accounting for the
  lepton mixing angles the CP asymmetry in right-handed neutrino
  decays vanishes in the limit in which the mixing pattern is
  exact. We study the implications that additional degrees of freedom
  from type II seesaw may have for leptogenesis in such a limit. We
  classify in a model independent way the possible realizations of
  type I and II seesaw schemes, differentiating between classes in
  which leptogenesis is viable or not. We point out that even with the
  interplay of type I and II seesaws there are generic classes of
  minimal models in which the CP asymmetry vanishes. Finally we
  analyze the generation of the lepton asymmetry by solving the
  corresponding kinetic equations in the general case of a mild
  hierarchy between the light right-handed neutrino and the scalar
  triplet masses. We identify the possible scenarios in which
  leptogenesis can take place.
\end{abstract}
\end{titlepage}
\setcounter{footnote}{0}
\section{Introduction}
\label{sec:intro}
Leptogenesis is a scenario in which the baryon asymmetry of the
Universe is generated first in the lepton sector and partially
reprocessed into a baryon asymmetry via standard model electroweak
sphaleron processes. Models for Majorana neutrino masses usually
involve new interactions that satisfy the necessary conditions for
leptogenesis namely lepton number violation, CP violation in the
lepton sector and departure from thermodynamic equilibrium (provided
by the expansion of the Universe). Type-I seesaw \cite{seesaw} is
certainly the most well studied framework for leptogenesis, however
further analysis in other Majorana neutrino mass models such as
type II \cite{Schechter:1980gr} and type III \cite{Foot:1988aq}
seesaws has also been carried out (see ref. \cite{Davidson:2008bu} for
more details).

Even in the light of the most recent neutrino data
\cite{Fogli:2011qn, Schwetz:2011zk} there is still a strong motivation to believe
the mixings in the lepton sector are governed by an underlying flavor
symmetry (for a review see e.g. \cite{Altarelli:2010gt} and references
therein). Indeed it is in part due to this observation that
leptogenesis in Majorana neutrino mass models has been recently
considered. In particular, there is an extensive literature in the
case of type I seesaw
\cite{Jenkins:2008rb,Hagedorn:2009jy,Bertuzzo:2009im,
  AristizabalSierra:2009ex,Felipe:2009rr} in which it has been shown
that as long as the flavor symmetry enforces a concrete flavor mixing
scheme (e.g. tribimaximal (TB) mixing) the CP violating asymmetry
$\epsilon_N$ vanishes in the limit in which such a pattern is exact\footnote{All 
  these analyses have been carried out assuming
  leptogenesis takes place in the flavor broken phase. Different
  results may be found if the generation of the $B-L$ asymmetry occurs
  in the flavor symmetric phase
  \cite{AristizabalSierra:2007ur,AristizabalSierra:2009bh,Sierra:2011vk}.}.

Viable leptogenesis requires either departures from the exact mixing
scheme, or the presence of extra degrees of freedom beyond those
present in the standard type I seesaw (three electroweak singlet
right-handed (RH) neutrinos). Departures could arise for example via next-to-leading order corrections (${\cal
  O}(v_F^2/\Lambda^2)$ corrections) where $v_F$ is the characteristic
scale at which the flavor symmetry is broken and $\Lambda$ is a cutoff
scale, or as recently proposed in \cite{Cooper:2011rh} through renormalisation group corrections.

In ref. \cite{AristizabalSierra:2009ex}
it has been pointed out that interplay between type I and type II seesaws
may allow leptogenesis at leading order (${\cal
  O}(v_F/\Lambda)$). It is the purpose of this paper to study under
which conditions such scenarios are actually obtained from the inclusion of extra degrees of freedom arising
from type II seesaw (involving one or more electroweak triplet scalars).

The interplay between type I and II seesaws for neutrino masses is an
almost unavoidable feature of left-right symmetric models. It has been
thoroughly considered in ref. \cite{Akhmedov:2006de}
without assuming any underlying flavor symmetry. As a framework for
leptogenesis it has been analysed in references
\cite{Hambye:2003ka,Hambye:2003rt, Antusch:2004xy, Hambye:2005tk, Antusch:2007km}, showing that the differences with the standard leptogenesis scenario can be
striking. This interplay has also been considered in flavor models
(see e.g. \cite{Chen:2005jm}). We extend upon these analyses by
exploring the feasibility of leptogenesis in ``hybrid'' type I +
type II flavor models aiming to differentiate
between those that might allow leptogenesis to proceed and
those in which even with the presence of the new degrees of freedom
the viability of leptogenesis relies on the departures from the exact mixing scheme (as is the case when only type I seesaw is present).

The rest of the paper is organized as follows: in section
\ref{sec:exact} we set up our notation and work out the implications a
flavor symmetry has in mixed I+II schemes by classifying
the different possible realizations. In section
\ref{sec:Leptogenesis-in-the-mixed-scheme} we elaborate on the
implications for leptogenesis in the different realizations found in
section \ref{sec:exact} whereas in section \ref{sec:BmLasymmetry} we
work out the calculation of the lepton asymmetry in scenarios in which
the hierarchy between the lightest RH neutrino and the electroweak
triplet is mild. We present our conclusions and final remarks in
section \ref{sec:conclusions}. In appendix~\ref{sec:appendix} we summarize
the formulas used in the calculation of the lepton asymmetry.
\section{Exact mixing schemes}
\label{sec:exact}
The presence of electroweak singlet RH neutrinos and a scalar $SU(2)$
triplet is described by the following Lagrangian
\begin{equation}
  \label{eq:full-L}
  {\cal L} = {\cal L}^{I} +{\cal L}^{II}\,,
\end{equation}
separating the type I and type II seesaw Lagrangians
that, in the basis in which the RH neutrinos Majorana mass matrix is
diagonal, can be written as
\begin{align}
  \label{eq:Lag-I}
  {\cal L}^{I}=& -\lambda_{ij}\overline N_{R_i} \ell_{L_j}
  \tilde H^\dagger
  -\frac{1}{2}\overline N_{R_i}C M_{R_i}
  \overline N_{R_i}^T + \mbox{h.c.}\,,
  \\
  \label{eq:Lag-II}
  {\cal L}^{II}=&-Y_{ij}\ell_{L_i}^TCi\tau_2\pmb{\Delta}\ell_{L_j}
  -M^2_\Delta \mbox{Tr}\pmb{\Delta}^\dagger\pmb{\Delta}
  +\mu H^Ti\tau_2\pmb{\Delta} H + \mbox{h.c.}.
\end{align}
Where we left the standard $H$ Higgs doublet potential implicit. Here $\ell_L=(\nu_L, l_l)^T$ and $H=(h^+, h^0)^T$ are the lepton and
Higgs $SU(2)$ doublets ($\widetilde H=i\tau_2 H^*$), $N_{R_i}$ are the
RH neutrinos, $C$ is the charge conjugation operator, $\pmb{\lambda}$
and $\pmb{Y}$ are matrices in flavor space and $\pmb{\Delta}$, the $SU(2)$
scalar electroweak triplet \footnote{The generalisation for multiple triplets is straightforward and considered implicitly in the following sections whenever appropriate.}  has hypercharge +1 (to the lepton doublets -1/2) and is given by
\begin{equation}
  \label{eq:triplet}
  \pmb{\Delta}=
  \begin{pmatrix}
    \Delta^{++} & \Delta^{+}/\sqrt{2}\\
    \Delta^{+}/\sqrt{2} & \Delta^{0}
  \end{pmatrix}\,.
\end{equation}
After electroweak symmetry breaking the setup of eq. (\ref{eq:Lag-I})
and (\ref{eq:Lag-II}) leads to the following effective light neutrino
mass matrix:
\begin{equation}
  \label{eq:eff-mass-matrix}
  \pmb{m_\nu^{\text{eff}}}=\pmb{m_\nu^I} + \pmb{m_\nu^{II}}=-\pmb{m_D}^T\pmb{\hat M_R}^{-1}\pmb{m_D} + 2\,v_\Delta\,\pmb{Y}\,,
\end{equation}
where the first term is the contribution from type I whereas the
second one is the type II contribution. The Dirac mass matrix is
defined as usual $\pmb{m_D}=v\pmb{\lambda}$ (with $v\simeq 174$ GeV),
and given the interactions in eq. (\ref{eq:Lag-II}) the triplet vacuum
expectation value can be written as
$\langle\Delta^0\rangle=v_\Delta=\mu^*\,v^2/M_\Delta^2$. For the
following discussion it is useful to write the type I seesaw
contribution as a tensor product of the parameter space vectors
$\pmb{m_{D_i}}^T=\left(m_{D_{i1}},m_{D_{i2}},m_{D_{i3}}\right)$:
\begin{equation}
  \label{eq:nmmI-recasted}
  \pmb{m_\nu^I}= - \sum_{i=1}^N M_{N_i}^{-1}\,\pmb{m_{D_i}}\otimes \pmb{m_{D_i}}\,,
\end{equation}
with $N$ determined by the number of RH electroweak singlet neutrinos and $\pmb{m_{D_i}}^T$ having the same number of rows.
 
Our main assumption is that the Lagrangian in eq. (\ref{eq:full-L}) originates from
an underlying Lagrangian invariant under a $G_F$ flavor group that gets broken and enforces a concrete lepton
mixing scheme in which the light neutrino mass matrix and the
matrices that define it are form-diagonalizable \cite{Low:2003dz}.
For concreteness hereafter we will
consider the TB scheme, but we do this without loss of generality as our analysis remains valid 
regardless of the mixing scheme (provided $G_F$ guarantees
$\pmb{m_\nu^\text{eff}}$ to be form-diagonalizable). Due to our
assumption and mixing pattern choice the effective light neutrino mass
matrix becomes
\begin{equation}
  \label{eq:tbm-light-mm}
    \pmb{m_\nu^{\text{eff}}}=\pmb{m_\nu^I} + \pmb{m_\nu^{II}}
    = 
    \begin{pmatrix}
      4a + b & -2a +b & -2a + b\\
      \cdot & a+b+c & a+b-c\\
      \cdot & \cdot & a+b+c
    \end{pmatrix}\,.
\end{equation}
This matrix is diagonalized by the $\pmb{U_\text{TB}}$ leptonic mixing
matrix, namely
\begin{equation}
  \label{eq:diagonalization}
  \pmb{U_\text{TB}}^T\,\pmb{m_\nu^\text{eff}}\,\pmb{U_\text{TB}}=
  \pmb{\hat m_\nu^\text{eff}}=\text{diag}(6a,3b,2c)=
  \text{diag}(m_{\nu_1},m_{\nu_2},m_{\nu_3})\,,
\end{equation}
\begin{equation}
  \label{eq:UTB}
  \pmb{U_\text{TB}}=(\pmb{v_1},\pmb{v_2},\pmb{v_3})=
  \begin{pmatrix}
    \sqrt{2/3}  & 1/\sqrt{3} & 0\\
    -1/\sqrt{6} & 1/\sqrt{3} & -1/\sqrt{2}\\
    -1/\sqrt{6} & 1/\sqrt{3} & 1/\sqrt{2}
  \end{pmatrix}\,.
\end{equation}
$\pmb{v_i}$ are the eigenvectors of
$\pmb{m_\nu^\text{eff}}$
($\pmb{m_\nu^\text{eff}}\,\pmb{v_i}=m_{\nu_i}\,\pmb{v_i}$). Using
the eigenvector decomposition of the TB leptonic mixing matrix
in~(\ref{eq:UTB}) and the diagonalization relation
(\ref{eq:diagonalization}), the effective neutrino mass matrix can be
written as an outer product of the these characteristic vectors
\footnote{This decomposition can be understood as a consequence
  of $\pmb{m_\nu^\text{eff}}$ being form-diagonalizable.}:
\begin{equation}
  \label{eq:outer-product}
  \pmb{m_\nu^\text{eff}} = \sum_{i=1,2,3}m_{\nu_i} 
  \pmb{v_i}\otimes\pmb{v_i}\,,
\end{equation}
meaning the full light neutrino matrix is built-up from the sum of 3 matrices
each arising from the respective outer product.
Equations (\ref{eq:tbm-light-mm}) and (\ref{eq:diagonalization}) strongly imply
that the seesaw I and II mass matrices $\pmb{m_{\nu}^{I,II}}$ are
both diagonalized by $\pmb{U_{\text{TB}}}$.
Strictly this needs not be the case, but if not then we are in a situation where the two separate mechanisms conspire to cancel the TB incompatible contribution, and given their separate origins this is in principle quite unnatural (note that being diagonalised by $\pmb{U_{\text{TB}}}$ does not mean that $\pmb{U_{\text{TB}}}$ is the only matrix that diagonalises $\pmb{m_{\nu}^{I,II}}$, due to possible degeneracy in eigenvalues). We therefore safely assume these matrices consist of the same eigenvectors $\pmb{v_i}$ from which
$\pmb{m_\nu^\text{eff}}$ is constructed i.e.
\begin{equation}
  \label{eq:sesaw-mm-outer}
    \pmb{m_\nu^X} = \sum_{i=1,2,3}m_{\nu_i}^X 
  \pmb{v_i}\otimes\pmb{v_i}\qquad (X=I,II)\,.
\end{equation}
Although there are different ways given contributions within the same mechanism produce these outer products and lead to TB we can always express them in this way (it may require non-independent or vanishing $m_{\nu_i}^X$). In any case, the TB structure of $\pmb{m_\nu^\text{eff}}$ can arise in different ways which we classify according to the different possible structures of the matrices $\pmb{m_\nu^{I,II}}$. One main distinction is whether the identity matrix $\mathds{I}$ is present in one or both of the matrices. This is always possible as a contribution proportional to the identity does not alter the eigenstates, it simply shifts all the eigenvalues.
Three generic cases can be identified:
\begin{enumerate}[i.]
\item \label{case1} In the most general case $\pmb{m_\nu^I}$ as
  well as $\pmb{m_\nu^{II}}$ are entirely determined by all three
  eigenvectors $\pmb{v_i}$, with different parameters entering in each
  of all them, namely
\begin{equation}
  \label{eq:caseA}
  \pmb{m_\nu^\text{eff}}=\sum_{\substack{X=I,II\\i=1,2,3}}m_{\nu_i}^X\,
  \pmb{v_i}\otimes \pmb{v_i}\,.
\end{equation}
Within the general case parameters may vanish so that not all eigenvectors are repeated across both seesaws. When several parameters vanish it is useful to consider that situation as a separate case explicitly as below.
\item \label{case2} One special case corresponds to $\pmb{m_\nu^I}$
  ($\pmb{m_\nu^{II}}$) being proportional to the identity matrix. Here
  of course we deal with two options as follows:
\begin{align}
  \label{eq:different-cases-C}
  \pmb{m_\nu^I}&=\tilde m_\nu^I\,\mathds{I}
  \quad\text{and}\quad
  \pmb{m_\nu^{II}}=\sum_{i=1,2,3} m_{\nu_i}^{II}\,\pmb{v_i}\otimes \pmb{v_i}\,,\\
  \pmb{m_\nu^I}&=\sum_{i=1,2,3} m_{\nu_i}^{I}\,\pmb{v_i}\otimes \pmb{v_i}
  \quad\text{and}\quad
  \pmb{m_\nu^{II}}=\tilde m_\nu^{II}\,\mathds{I}\,.
\end{align}
\item \label{case3} Another special case has
  $\pmb{m_\nu^I}$ ($\pmb{m_\nu^{II}}$) arise from a single
  $\pmb{v_i}$ whereas $\pmb{m_\nu^{II}}$
  ($\pmb{m_\nu^I}$) arises without that $\pmb{v_i}$. As there are 3
  eigenvectors there are 3+3 options:
  \begin{align}
    \label{eq:different-cases-caseB1}
    \pmb{m_\nu^I}&=\sum_{i=1,2}^j m_{\nu_i}^I\,\pmb{v_i}\otimes \pmb{v_i}\quad
     (i<j=2,3)\quad ; \quad
     \pmb{m_\nu^{II}}=m_{\nu_k}^{II}\,\pmb{v_k}\otimes \pmb{v_k}\quad (k\neq i)\,,
     \\
     \label{eq:different-cases-caseB2}
     \pmb{m_\nu^{I}}&=m_{\nu_i}^I\,\pmb{v_i}\otimes \pmb{v_i}\quad (i\neq k)
     \quad ; \quad
     \pmb{m_\nu^{II}}=\sum_{k=1,2}^j m_{\nu_k}^{II}\,\pmb{v_k}\otimes \pmb{v_k}\quad
     (k<j=2,3)\,.
  \end{align}
\end{enumerate}

These cases can be further distinguished by the number of parameters
defining the full light neutrino mass matrix. Depending on the presence of $\mathds{I}$ we may have up to two extra parameters. Case \ref{case1} is
determined by up to 8 parameters, both options in \ref{case2} either have 3 or 4 and the
options in \ref{case3} by 2 to 5. In that sense models of type
\ref{case3} can be regarded as the minimal realizations of I+II mixed
schemes with an underlying flavor symmetry accounting for the exact mixing scheme. From now on we will refer to them as ``minimal models''. We
adopt the notation (a/b) denoting how many eigenstates are present in each seesaw type as follows: in the most minimal of cases we have (1/1) as the extreme of either (2/1) for those of type
(\ref{eq:different-cases-caseB1}) and (1/2) referring to the ones in
(\ref{eq:different-cases-caseB2}). General models instead are
labeled as (3/3) while the intermediate models in \ref{case2} would be either ($\mathds{I}$/3) or (3/$\mathds{I}$), or ($\mathds{I}$/2) or (2/$\mathds{I}$) in their respective extreme (whereas ($\mathds{I}$/1) and (1/$\mathds{I}$) are not considered as they only determine one eigenvector).
\subsection{General models}
\label{sec:general-mod}
From eqs. (\ref{eq:eff-mass-matrix}), (\ref{eq:nmmI-recasted}) and
(\ref{eq:caseA}) we can determine the structure of the Dirac matrix as well as the
structure of the Yukawa matrix $\pmb{Y}$ involved in type II seesaw:
\begin{align}
  \label{eq:33models1}
  \pmb{m_D}&=(\tilde \lambda_1 \pmb{v_1},\tilde \lambda_2 \pmb{v_2},
  \tilde \lambda_3 \pmb{v_3})^T\,,\\
  \label{eq:33models2}
  \pmb{Y}&=\sum_{k=1,2,3} \tilde y_k\,\pmb{v_k}\otimes \pmb{v_k}\,.
\end{align}
Note that we made use of the fact that the parameter space
vectors $\pmb{m_{D_i}}$ and the Yukawa matrix $\pmb{Y}$ are defined by
the eigenvectors $\pmb{v_i}$ up to global factors $\tilde \lambda_i$
and $\tilde y_i$.  The case above is the most general type of model, which we denoted as (3/3) and they are defined in the
type I sector by three RH neutrinos. In this case all 3 eigenvectors are explicitly present in both seesaws.
\subsection{Intermediate models}
\label{sec:interm-models}

In this class of models, the exact mixing
structure of the full effective light neutrino mass matrix arises from
either the type I sector ((3/$\mathds{I}$) and (2/$\mathds{I}$) models) or the type II sector
(($\mathds{I}$/3) and ($\mathds{I}$/2) models). The sector proportional to the identity matrix just
adds a parameter shifting the eigenvalues. Accordingly, the structures of the
Dirac mass matrix and $\pmb{Y}$ are given by:
\begin{align}
  \label{eq:intermmodels1}
  \pmb{m_D}&=(\tilde \lambda_1 \pmb{v_1},\tilde \lambda_2 \pmb{v_2},
  \tilde \lambda_3 \pmb{v_3})^T,&
  \pmb{Y}&=\tilde y\, \mathds{I},\\
  \label{eq:intermmodels2}
  \pmb{m_D}&=\tilde \lambda\, \mathds{I},&
  \pmb{Y}&=\sum_{k=1,2,3} \tilde y_k\,\pmb{v_k}\otimes \pmb{v_k}\,,
\end{align}
with obvious generalisation if only 2 eigenvectors are explicitly present.
\subsection{Minimal (2/1) models}
\label{sec:minimal-21-models}
Comparing eq. (\ref{eq:nmmI-recasted}) with
eq. (\ref{eq:different-cases-caseB1}) (left-hand side) it becomes clear that minimal (2/1) models correspond in the type I sector to
minimal type I seesaw models (with only two RH neutrinos).
Moreover from these equations it can be seen the Dirac mass matrix is
given by
\begin{equation}
  \label{eq:dirac-mm-21models}
  \pmb{m_D}=v(\tilde \lambda_i\,\pmb{v_i},\tilde \lambda_j\,\pmb{v_j})^T\,,\qquad
  i\neq j\,.
\end{equation}
The type II sector Yukawa matrix $\pmb{Y}$ can be
determined from the corresponding mass matrix in
(\ref{eq:eff-mass-matrix}) and eq. (\ref{eq:different-cases-caseB1})
(right-hand side):
\begin{equation}
  \label{eq:Y-21models}
  \pmb{Y}=\tilde y_k\,\pmb{v_k}\otimes \pmb{v_k}\,,
  \qquad k\neq i, j\,.
\end{equation}
We note that particular realizations of the (2/1) models are not strictly minimal if we do not require both seesaws to be present. What we could denote as (2/0) models by eliminating type II seesaw a massless eigenstate $v_k$ is possible (minimal type I with only two neutrinos). This is not viable for $k=2$ due to the observed squared mass splittings, but $k=1$ is consistent with a normal hierarchy and $k=3$ with an inverted hierarchy. In such a situation type I seesaw determines 2 eigenstates explicitly with the one corresponding to the vanishing eigenvalue having to be orthogonal \footnote{This is similar to what happens in sequential dominance scenarios where the third eigenvalue is approximately zero, see e.g. \cite{King:2005bj, deMedeirosVarzielas:2005ax}.}. Finally a particular case (of this and the (1/2) case) was already denoted as (1/1) and once again the solar eigenstate must explicitly arise from one or other type of seesaw due to the observed squared mass splittings.

\subsection{Minimal (1/2) models}
\label{sec:minimal-12-models}
From eq. (\ref{eq:nmmI-recasted}) and
(\ref{eq:different-cases-caseB2}) (left-hand side) we can see
(1/2) models have a single RH neutrino in the type I seesaw
sector. The Dirac mass matrix is therefore given by
\begin{equation}
  \label{eq:dirac-case12}
  \pmb{m_D}=v\,\tilde \lambda_i\,\pmb{v_i}^T\,.
\end{equation}
In the type II sector the Yukawa matrix $\pmb{Y}$ is determined
by
\begin{equation}
  \label{eq:yukawa-case12}
  \pmb{Y}=\tilde y_j\,\pmb{v_j}\otimes\pmb{v_j} + \tilde y_k\,\pmb{v_k}\otimes\pmb{v_k}\,,
  \qquad j<k\,,\qquad i\neq j,k\,.
\end{equation}
In contrast to the (2/1) models in this case the light
neutrino spectrum arises from a non-vanishing eigenvalue from I and
two non-vanishing eigenvalues from the type II contribution. Analogously, dropping type I seesaw could result in viable models with a massless eigenstate.
When considering this or other cases with more than one eigenvector represented in type II it is relevant to consider if multiple $\Delta^a$ have specific $\pmb{Y}^a$ structures associated with them that differ from the ones that can be constructed directly from the eigenvectors. It is possible and even natural to have the underlying symmetry specify the structures differently despite resulting in the exact mixing - in our parametrisation this is equivalent to having specific relations between the $m_{\nu_i}^{II}$ (see eq.(\ref{eq:sesaw-mm-outer})). It may happen then that there is a single $\Delta$ and more than one non-zero (but related) $m_{\nu_i}^{II}$.
\section{Leptogenesis in mixed schemes}
\label{sec:Leptogenesis-in-the-mixed-scheme}
Leptogenesis in mixed type I and type II general models has been
analysed in ref. \cite{Hambye:2003ka,Hambye:2003rt,Antusch:2004xy, Hambye:2005tk, Antusch:2007km}.  In
what follows we will present the standard formulas for the CP
violating asymmetry that we then particularize to the classes
discussed in the previous section. We also analyse the generation
of the lepton asymmetry in the case in which both the dynamics of the
lightest RH neutrino as well as the dynamics of the scalar triplet
have to be taken into account. Our discussion will be done in the
unflavored regime but can be readily extended to the flavored case.
\subsection{The CP violating asymmetries}
\label{sec:epsilon}
In models with RH neutrinos and scalar $SU(2)$ triplets the lepton
asymmetry can arise via the CP violating and out-of-equilibrium decays
of the RH neutrinos, the triplets or both. The CP violating asymmetry
in RH neutrino decays arises from the interference of the tree-level
process $N_i\to \ell\,\tilde H^\dagger$ and standard one-loop
corrections of the wave-function and vertex type. Due to the trilinear
scalar coupling in (\ref{eq:Lag-II}) there is in addition a new
one-loop vertex correction with the electroweak triplet scalar flowing
in the loop. The calculation of these
interferences yields \footnote{Eq. (\ref{eq:RHN-epsilon2}) follows from \cite{Antusch:2004xy} which differs from \cite{Hambye:2003ka} by a factor of $3/2$.} \cite{Hambye:2003ka,Antusch:2004xy}
\begin{align}
  \label{eq:RHN-epsilon1}
  \epsilon_{N_k}&=\frac{1}{8\,\pi\,v^2}
  \frac{1}{\left(\pmb{m_D}\,\pmb{m_D}^\dagger\right)_{kk}}
  \sum_{j\neq k} \mathfrak{I}\mbox{m} 
  \left[ 
    \left(
      \pmb{m_D}\pmb{\,m_D}^\dagger 
    \right)^2_{kj}
  \right]\,f(z_k)\,,\\
  \label{eq:RHN-epsilon2}
  \epsilon_{N_k}^\Delta&=-\frac{3}{8\,\pi\,M_\Delta}
  \frac{1}{
    \left(
      \pmb{m_D}\,\pmb{m_D}^\dagger
    \right)_{kk}}
  \mathfrak{I}\mbox{m}
  \left[
    \left(\pmb{m_D}\,\pmb{Y}^*\pmb{m_D}^T
    \right)_{kk}\,\mu
  \right]\,g(\omega_k)\,.
\end{align}
The functions $f(z_k)$ and $g(\omega_k)$, with $z_k=M_{N_j}^2/M_{N_k}^2$
and $\omega_k=M_\Delta^2/M_{N_k}^2$, are loop functions given by
\begin{align}
  \label{eq:loop-functions}
  f(z_k)&=\sqrt{z_k}
  \left[
    \frac{2-z_k}{1-z_k} - (1+z_k)
    \log
    \left(
      \frac{1+z_k}{z_k}
    \right)
  \right]\,,\\
  g(\omega_k)&=\sqrt{\omega_k}
  \left[
    1 - \omega_k\log
    \left(\frac{1+\omega_k}{\omega_k}\right)
  \right]\,.
\end{align}
These functions have different limits depending on the RH neutrino
mass spectrum and on the hierarchy between the RH neutrinos and
$SU(2)$ triplet scalar mass. In the case of a hierarchical RH
neutrino spectrum the function $f(z_k)$ can be accurately approximated
as
\begin{equation}
  \label{eq:fzk-approx}
  f(z_k)\to -\frac{3}{2\,\sqrt{z_k}}\,,
\end{equation}
whereas the function $g(\omega_k)$ according to
\begin{align}
  \label{eq:gomegak-approx}
  g(\omega_k)\simeq\sqrt{\omega_k}
  \quad\mbox{if}\quad \omega_k\ll 1\quad\mbox{or}\quad
  g(\omega_k)\simeq\frac{1}{2\,\sqrt{\omega_k}}
  \quad\mbox{if}\quad \omega_k\gg 1\,.
\end{align}
The CP violating asymmetry in $\pmb{\Delta}$ decays involves only the
interference between the tree-level process $\pmb{\Delta}\to
\ell\,\ell$ and a vertex one-loop diagram induced again by the
trilinear scalar coupling. The result reads \cite{Hambye:2003ka}
\begin{equation}
  \label{eq:triplet-CPasymmetry}
  \epsilon_\Delta=-\frac{1}{8\,\pi\,v^2}\frac{1}{M_\Delta}
  \frac{
    \sum_k\mathfrak{I}\mbox{m}
    \left[
      \left(
        \pmb{m_D}\,\pmb{Y}^*\,\pmb{m_D}^T
      \right)_{kk}\,\mu
    \right]}{\mbox{Tr}\left[\pmb{Y}\,\pmb{Y}^\dagger\right] + \mu^2/M_\Delta^2}\,
  H(\omega_k)\,,
\end{equation}
where the loop function in this case is given by
\begin{equation}
  \label{eq:loop-function-triplet-decay}
  H(\omega_k)=\frac{1}{\sqrt{\omega_k}}\log\left(1 + \omega_k \right)\,.
\end{equation}
In the limits $\omega_k\ll 1$ and $\omega_k\gg 1$ this
function can be respectively written as
\begin{equation}
  \label{eq:homegak-limit}
  H(\omega_k)\simeq 1-\frac{\omega_k}{2}\qquad\mbox{and}\qquad
  H(\omega_k)\simeq\frac{\log\omega_k}{\omega_k}\,.
\end{equation}

With the results in (\ref{eq:RHN-epsilon1}), (\ref{eq:RHN-epsilon2})
and (\ref{eq:triplet-CPasymmetry}) at hand we are now in the position
to study the viability of leptogenesis in the models discussed in
sec. \ref{sec:exact}. We start by noting
that $\epsilon_{N_k}$ is simply the CP violating
asymmetry governing standard (type I) leptogenesis, so the results from
refs. \cite{Jenkins:2008rb,Hagedorn:2009jy,Bertuzzo:2009im,
  AristizabalSierra:2009ex,Felipe:2009rr} apply
i.e. $\epsilon_{N_k}=0$. This means that if leptogenesis is to be
possible its feasibility must rely on either a non-vanishing
$\epsilon_{N_k}^\Delta$ or $\epsilon_\Delta$. Therefore, the relevant
expression for analysis is the matrix
\begin{equation}
  \label{eq:relevant-matrix}
  \pmb{{\cal M}}=\pmb{m_D}\pmb{Y}^*\pmb{m_D}^T\,.
\end{equation}
Since in general $G_F$ does not constrain $\mu$ to be real this
implies viable leptogenesis is possible even if $\pmb{{\cal M}}$ is
real.

In the case of general (3/3) models, given the structures for
$\pmb{m_D}$ and $\pmb{Y}$ in (\ref{eq:33models1}) and
(\ref{eq:33models2}), the relevant quantity $\pmb{{\cal M}}$ reads
\begin{equation}
  \label{eq:M-for-33models}
  \pmb{\cal M}=\mbox{diag}(\tilde \lambda_1^2\,
  \tilde y_1^*, \tilde \lambda_2^2\, \tilde y_2^*,\tilde \lambda_3^2\,
  \tilde y_3^*)\,,
\end{equation}
thus yielding a non-vanishing CP asymmetry and in principle
viable leptogenesis. These models represent a class of generic models
in which the presence of a flavor symmetry allows for leptogenesis
even at leading order in $v_F/\Lambda$ despite form-diagonalizable structures. This is to be compared with
the standard leptogenesis scenario in which a non-vanishing CP
asymmetry is possible only by the inclusion of next-to-leading order
corrections. 

\begin{figure}
  \centering
  \includegraphics[width=9cm,height=6cm]{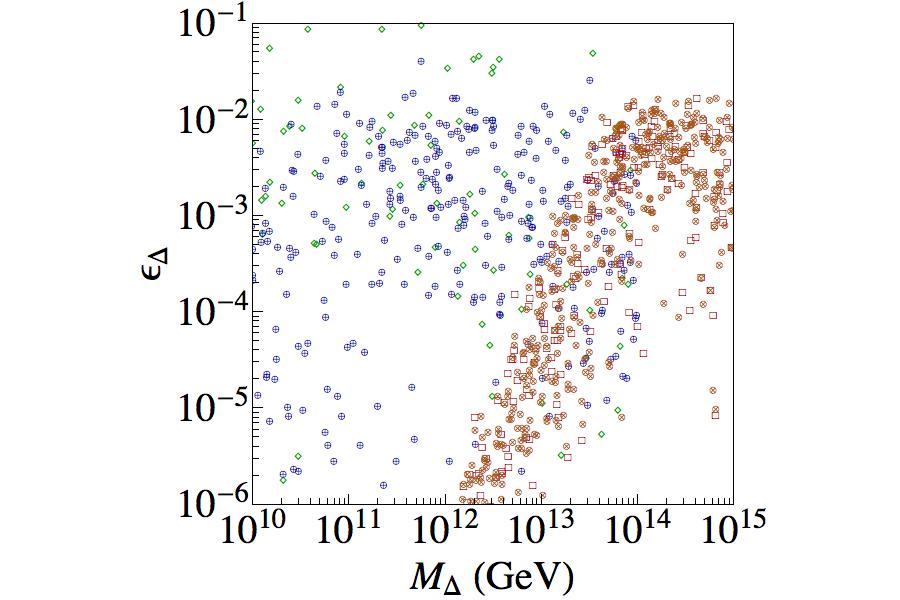}
  \caption{\it $\epsilon_\Delta$ as function of $M_\Delta$. Red squares and orange crosses for normal and inverted hierarchy of a specific 3-parameter predictive case. Green squares and blue crosses for normal and inverted hierarchy of the general 8-parameter case.}
  \label{fig1}
\end{figure}

In general cases it is not possible to do predictions on the asymmetry because we have up to 8 complex parameters to fit 3 physical mass quantities and 2 Majorana phases. Therefore it is interesting to consider predictive scenarios defined in the following way:
\begin{itemize}
\item[$a)$] the masses $m_{\nu_j}$ are described by a total of at most 3 complex parameters (at least 2 to fit the mass splitting). Let us call these parameters as $a_i^X$, with again $X=I,II$;
\item[$b)$] type-I and type-II seesaw give independent contributions. For example if type-I gives an $\mathds{I}$ contribution, type-II does not;
\item[$c)$] the masses are linear in the parameters $a_k^X$, that is  $m_{\nu_j}^X= m_{\nu_j}^X( a_k^X)$.
\end{itemize}
According to the previous statements it is clear that these conditions for predictivity apply to either the intermediate or minimal classes from sec. \ref{sec:exact}, for which we can  study connections between high and low energy CP violation parameters.
Before moving on from the general case we explore further the predictive conditions as defined in $a)$, $b)$ and $c)$.
Consider that for both seesaw types the functions defined at point $c)$  have the form  
\beq
\label{mseq}
m_{\nu_j}^X= a_0^X +a_j^X \,,
\eeq
with $a_0^X$ a term that is present for cases with $\mathds{I}$ and vanishes otherwise.
The $\epsilon_\Delta$ asymmetry then depends on a combination of the parameters from eq. (\ref{mseq})
\begin{equation}
\label{explicit}
\mbox{Im} \left[ (a_0^I a_0^{II \star}) + \sum_i \left( a_i^I a_i^{II \star} \right) + \left(\sum_i a_i^I \right) a_0^{II \star} + a_0^I \left(\sum_i a_i^{II \star}\right) \right]\,.
\end{equation}
The $\epsilon^\Delta_{N_i}$ dependences are very similar but without the sums over $i$. Technically it would be possible for $\epsilon_\Delta$ to vanish and $\epsilon^\Delta_{N_i}$ to be non-zero, but this requires a specific relationship between the $a_i^X$ parameters.
The general case with 8 parameters (3 $a_i^I$, 3 $a_i^{II}$, $a_0^I$ and $a_0^{II}$) is illustrated in fig. \ref{fig1} (green squares and blue crosses for normal and inverted hierarchy respectively).

We consider now intermediate models of type (3/$\mathds{I}$) or ($\mathds{I}$/3). The explicit form of
$\pmb{{\cal M}}$ can be calculated from (\ref{eq:intermmodels1})
and (\ref{eq:intermmodels2}), the result reads
\begin{align}
  \label{eq:M-for-I3-models}
  \pmb{{\cal M}}&=\tilde y \,\mbox{diag}(\tilde\lambda_1^2,\tilde\lambda_2^2,
  \tilde\lambda_3^2)\,,\nonumber\\
  \pmb{{\cal M}}&=\frac{\tilde\lambda^2}{3}
  \begin{pmatrix}
    2 \tilde y_1 +\tilde y_2 & -\tilde y_1 + \tilde y_2 & -\tilde y_1 + \tilde y_2\\
    \cdot                & \frac{\tilde y_1}{2} + \tilde y_2 + \frac{3}{2}\tilde y_3 &
    \frac{\tilde y_1}{2} + \tilde y_2 - \frac{3}{2}\tilde y_3\\
    \cdot & \cdot & \frac{\tilde y_1}{2} + \tilde y_2 + \frac{3}{2}\tilde y_3
  \end{pmatrix}\,.
\end{align}
As we can see, just like in the general (3/3) models the CP asymmetries do not vanish and qualitatively all the conclusions derived in that case hold
in (3/$\mathds{I}$) and ($\mathds{I}$/3) models as well - this is consistent with \cite{Antusch:2005tu} where a type II seesaw contribution proportional to the identity was considered. The
question of whether leptogenesis is
feasible becomes a quantitative question.
The predictive cases include (2/$\mathds{I}$) and ($\mathds{I}$/2), and considering the particular form of eq. (\ref{explicit}) $\epsilon_\Delta$ becomes a function of
\beq
\mbox{Im}\left[(a_0^X)^\star \sum_i a_i^Y \right]\,,
\eeq
with $X=I,Y=II$ or vice versa and $i$ runs over two indices. The $\epsilon^\Delta_{N_i}$ depend on just one of the terms in the sum above.
Under each explicit case we can determine the mass differences in terms of the parameters, therefore constraining them. For (2/$\mathds{I}$) we have:
\bea
\Delta m_{\odot}^2 &=& \left| a^I_2 \right|^2 - \left| a^I_1 \right|^2 + 2 a_0^{II} \left( \left| a^I_2 \right| \cos \alpha_2^I - \left| a^I_1 \right| \cos \alpha_1^I \right)\\
\Delta m_{@I}^2&=& \left| a^I_2 \right|  \left( \left| a^I_2 \right| + 2 a_0^{II} \cos \alpha_2^I  \right)\\
\Delta m_{@N}^2&=& -\left| a^I_1 \right| \left( \left| a^I_1 \right| + 2 a_0^{II} \cos \alpha_1^I  \right) \,,
\eea
where $\Delta m_{\odot}^2$, $\Delta m_{@I}$ and $\Delta m_{@N}$ are the solar, atmospheric for inverted and atmospheric for normal hierarchy squared mass differences, $\alpha_i^I$ are the phases of the respective $a_i^I$ and we absorbed the phase of $a_0^{II}$ to make it real without loss of generality.
For illustration we present the results for the asymmetries of (2/$\mathds{I}$) in fig. \ref{fig1} (red squares and orange crosses for normal and inverted hierarchy respectively).

Similarly non-vanishing asymmetries can be generated in the 3 parameter classes (1+$\mathds{I}$/1) and (1/1+$\mathds{I}$), due to the presence of $\mathds{I}$ and therefore of a non-vanishing $a_0^X$.



We turn finally to the special case of the minimal models (2/1)
and (1/2) scenarios. For these the matrix $\pmb{{\cal M}}$ must vanish as a
consequence of the orthogonality of the $\pmb{v_i}$ eigenvectors due to $\pmb{Y}^a$ having the form
$\pmb{v_k}\otimes\pmb{v_k}$ and the absence of repeated eigenvectors.  This conclusion can be seen from
eqs. (\ref{eq:dirac-mm-21models}), (\ref{eq:Y-21models}) and
(\ref{eq:dirac-case12}), (\ref{eq:yukawa-case12}) respectively: in
either case we get two contributions: combining the single eigenvector
of one seesaw type with the other two of the other seesaw type.  The
two contributions to $\pmb{{\cal M}}$ necessarily share the matricial
form constructed of orthogonal eigenvectors $\pmb{v_i}^T . (\pmb{v_k}
\otimes \pmb{v_k})^\ast . \pmb{v_i}$ and individually vanish. This can
also be seen explicitly by considering eq. (\ref{explicit}) with $a_0^I
= a_0^{II}=0$ and no repeated eigenvectors meaning $\sum_i \left( a_i^I
  a_i^{II \star} \right)=0$ with all individual terms in the sum
vanishing (meaning $\epsilon_\Delta$ and all individual
$\epsilon^\Delta_{N_i}$ vanish).

This result may appear to be incompatible with the leptogenesis
discussion in \cite{deMedeirosVarzielas:2011tp} (where $\pmb{{\cal
    M}}$ does not vanish despite having exact TB mixing). There is in
fact no contradiction - the parametrisations used are both valid: here
we consider one quite useful for studying hybrid leptogenesis, relying
on the outer products $\pmb{v_k} \otimes
\pmb{v_k}$;\cite{deMedeirosVarzielas:2011tp} relies on structures that
simultaneously give insight into particular phenomenological
consequences and have direct links to particular symmetries. If one
takes a situation where the $\pmb{Y}^a$ are proportional to the
structures of \cite{deMedeirosVarzielas:2011tp} and re-express them
into our eigenvector parametrisation, we readily conclude that such
cases can lead to repeated eigenvectors, falling outside our minimal
classes and therefore are able to produce non-vanishing asymmetries
consistently with our analysis.

We conclude then that in these minimal classes as defined, even the
interplay between type I and type II seesaw does not enable
leptogenesis in the exact mixing limit. Leptogenesis becomes possible
only when deviations from that limit are introduced (for example via
the introduction of higher dimensional effective operators).

\begin{figure}
  \centering
  \includegraphics[width=7.4cm,height=5.6cm]{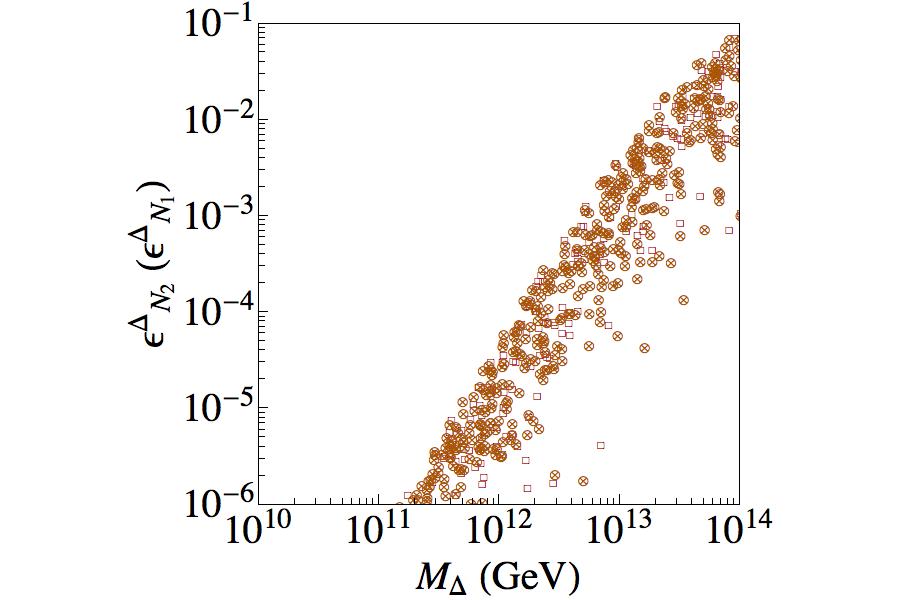}
  \includegraphics[width=7.4cm,height=5.6cm]{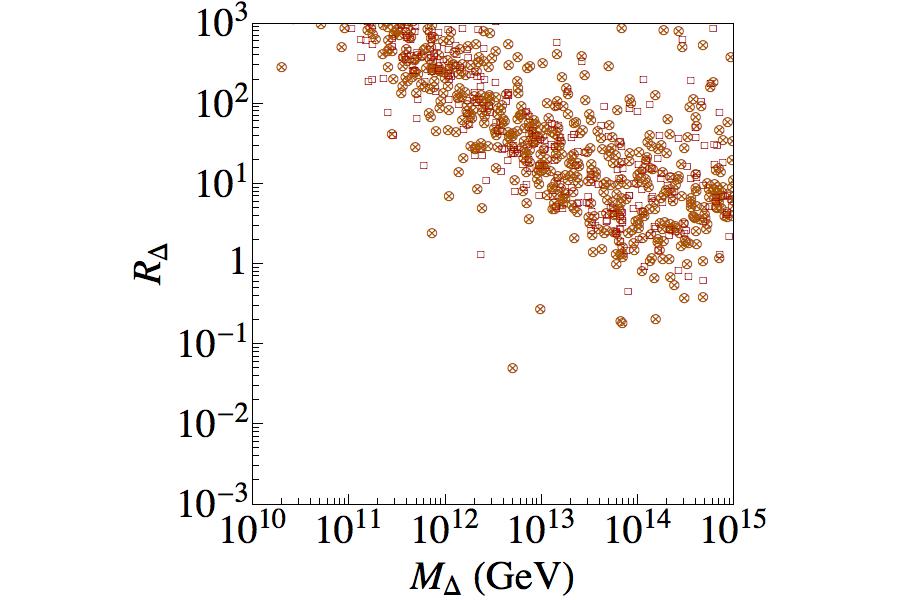}
  \caption{\it $\epsilon^\Delta_{N_i}$ (left-panel) and ratio $R_\Delta = \epsilon^\Delta_{N_i}/\epsilon_\Delta$
    (right-panel) as function of $M_\Delta$ for a specific 3-parameter
    predictive case.}
  \label{fig2}
\end{figure}

In fig. \ref{fig1} we compare a predictive case governed by 3
parameters in red squares and orange crosses (in this case a
(2/$\mathds{I}$)) against a general (3+$\mathds{I}$/3+$\mathds{I}$)
case governed by 8 parameters in green squares and blue crosses. In
fig. \ref{fig2} we
take the same (2/$\mathds{I}$) predictive case, present the values
for the relevant $\epsilon^\Delta_{N_i}$ and compare the relative by taking their ratio $R_\Delta = \epsilon^\Delta_{N_i}/\epsilon_\Delta$.

%

\subsection{Generation of the $\pmb{L}$ asymmetry}
\label{sec:BmLasymmetry}
In a mixed scheme as the one we are discussing here, and assuming a
hierarchical RH neutrino mass spectrum, leptogenesis can proceed in
different ways according to the hierarchy between the mass of the lightest RH neutrino, $M_{N_1}$ and
$M_\Delta$. If $M_{N_1}\ll M_\Delta$ the effects of $\pmb{\Delta}$ are
decoupled and the lepton asymmetry is generated via $N_1$
dynamics. Apart from the new contribution to the total CP asymmetry,
eq. (\ref{eq:RHN-epsilon2}), there is no difference between this case
and the standard leptogenesis scenario. Conversely, if $M_{N_1}\gg
M_\Delta$ the lepton asymmetry is entirely produced by the dynamics of
$\pmb{\Delta}$. The triplet having non-trivial standard model quantum
numbers couples to electroweak gauge bosons, so the determination of
the lepton asymmetry, even at leading order in the couplings $\pmb{Y}$,
involves gauge boson induced triplet annihilations
\cite{Hambye:2003rt,Hambye:2005tk}. One can also envisage a scenario
where $\pmb{\Delta}$ and $N_1$ being quasi-degenerate both generate the
lepton asymmetry and this is the scenario we want to discuss in what
follows.

In a hot plasma with $N$ lepton number and CP violating states
$S_1,\dots ,S_N$ the evolution of the lepton asymmetry
$L=\sum_{i=e,\mu,\tau}(2\ell_i+e_i)$ ($e_i$ being the RH charged
leptons) can be written, for a general mass spectrum, according to
\footnote{We do not include the change in the lepton densities due to
  sphaleron processes.}
\begin{equation}
  \label{eq:BmL-general}
  \dot Y_{\Delta_L}(z) = \sum_{i=1}^N\dot Y^{(S_i)}_{\Delta_L}(z)\,,
\end{equation}
where, following ref. \cite{Nardi:2007jp}, we are using the notation
$s(z)H(z)z \,d\,Y_X(z)/dz\equiv \dot
Y_X(z)$. Here $z=M_1/T$ ($M_1$ being the mass of the
lightest state), $Y_{\Delta_X}=n_X-n_{\bar X}/s$ with $n_X$ ($n_{\bar
  X}$) the number density of particles (antiparticles), $s$ the
entropy density and $H(z)$ the expansion rate of the Universe (the
definitions of these functions can be found in the appendix). $\dot Y_{\Delta_L}^{(S_i)}(z)$ is the asymmetry
generated by all the $N$ states. Note that in
(\ref{eq:BmL-general}) we have written the dimensionless inverse
temperature of the remaining states as $z_i=M_i/M_1\, z$.

In the case under consideration $N=2$ and the states correspond to the
lightest $SU(2)$ singlet neutrino and scalar triplet, therefore
\begin{equation}
  \label{eq:BmL-sinletplustriplet}
  \dot Y_{\Delta_L}(z)=\dot Y_{\Delta_L}^{(N_1)}(z)
  + \dot Y_{\Delta_L}^{(\Delta)}(z)\,.
\end{equation}
At leading order in the couplings $\pmb{\lambda}$ and $\pmb{Y}$ i.e.
${\cal O}(\lambda^2,Y^2)$ the two pieces are given by
\begin{align}
  \label{eq:BmL-1to2-sub}
  \dot Y_{\Delta_L}^{(N_1)}(z)&=\left(\dot
    Y_{\Delta_L}^{(N_1)}(z)\right)_{1\leftrightarrow 2} +
  \left(\dot
    Y_{\Delta_L}^{(N_1)}(z)\right)^{\text{sub}}_{2\leftrightarrow
    2}\,,
  \nonumber\\
  \dot Y_{\Delta_L}^{(\Delta)}(z)&=\left(\dot
    Y_{\Delta_L}^{(\Delta)}(z)\right)_{1\leftrightarrow 2} +
  \left(\dot
    Y_{\Delta_L}^{(\Delta)}(z)\right)^{\text{sub}}_{2\leftrightarrow
    2}\,.
\end{align}
The first terms in the right-hand side ($1\leftrightarrow 2$)
represent the contribution from the Yukawa induced decay and inverse
decay processes, $N_1\leftrightarrow \ell \tilde H^\dagger$ and
$\pmb{\Delta}\leftrightarrow \bar \ell\bar \ell$; the second terms
($2\leftrightarrow 2$) the off-shell Yukawa generated scattering
reactions $\ell \tilde H^\dagger\leftrightarrow \ell \tilde H^\dagger$
and $H^\dagger H^\dagger \leftrightarrow \ell\ell$, that even at
leading order must be included so to obtain kinetic equations with the
correct thermodynamical behaviour. 

In addition to the evolution of the $Y_{\Delta_L}$ asymmetry the full
network of Boltzmann equations should include the equations accounting
for the evolution of the RH neutrino and triplet number densities and
the triplet and Higgs asymmetries\footnote{These asymmetries are a
  consequence of these fields not being self-conjugate.}. A comment is
in order: the Higgs asymmetry in most of the studies of standard
leptogenesis is either neglected or treated as a spectator process
\cite{hep-ph/0512052}, in the case in question -or in the pure triplet
case- this in principle might be done if $\Gamma(\pmb{\Delta}\to
H\,H)\gg H(z)|_{z=1}$ can be guaranteed. This simplification, though
possible, is not necessary as one out of the five equations
governing the $L$ generation process can be removed by the condition
imposed by hypercharge neutrality \cite{Hambye:2005tk}:
\begin{equation}
  \label{eq:hypercharge-neutrality}
  2\,Y_{\Delta_\Delta} + Y_{\Delta H} - Y_{\Delta_L}=0\,.
\end{equation}
With the above considerations the relevant system of kinetic equations
can be written as
\begin{eqnarray}
  \label{eq:relevant-BEQs}
  \dot Y_{N_1}&=&-(y_{N_1}-1)\,\gamma_{D_{N_1}}\,,
  \nonumber\\
  \dot Y_\Sigma&=&-(y_\Sigma - 1)\,\gamma_{D_\Delta} 
  - 2(y_\Sigma^2 - 1)\,\gamma_A\,,\nonumber\\
  \dot Y_{\Delta_L}&=&\left[(y_{N_1}-1)\,\epsilon_{N_1}^\text{tot} -
    \left(y_{\Delta_L} + y_{\Delta_\Delta}^H\right)\right]\,\gamma_{D_{N_1}}\nonumber\\
  &&+\left[(y_\Sigma - 1)\,\epsilon_\Delta -
    2K_\ell\, (y_{\Delta_L} + y_{\Delta_\Delta})\right]\,\gamma_{D_\Delta}\,,
  \nonumber\\
  \dot Y_{\Delta_\Delta}&=&-\left[y_{\Delta_\Delta} + (K_\ell - K_H)\,y_{\Delta_L}
  + 2K_H \,y_{\Delta_\Delta}^H\right]\,,
\end{eqnarray}
where the following conventions have been adopted: $y_X\equiv
Y_X/Y_X^{\text{Eq}}$ (the exception being $y_{\Delta T}^H\equiv
Y_{\Delta_\Delta}/Y_H^\text{Eq}$ and
$y_{\Delta_L}=Y_{\Delta_L}/Y^\text{Eq}_\ell$),
$\Sigma\equiv\Delta + \Delta^\dagger$,
$\epsilon_{N_1}^\text{tot}\equiv\epsilon_{N_1}+\epsilon_{N_1}^\Delta$
and $\gamma_{D_{N_1}}$, $\gamma_{D_\Delta}$ and $\gamma_A$ are the
reaction densities for: RH neutrino and triplet decays and triplet
annihilations (expressions for these quantities are given in the
appendix).  The factors $K_{\ell,H}$ resemble the flavor projectors
defined in standard flavored leptogenesis
\cite{hep-ph/0601084,hep-ph/0605281} as they project triplet decays
into either the Higgs or the lepton doublet directions. They are
defined as follows
\begin{equation}
  \label{eq:projectors}
  K_\ell=\frac{\tilde m_\Delta^\ell}{\tilde m_\Delta^\ell + \frac{\tilde m_\Delta^2}
    {4\,\tilde m_\Delta^\ell}}\,
  \qquad\mbox{and}\qquad
  K_H=\frac{\tilde m_\Delta^2}{4\,\tilde m_\Delta^\ell\left(
    \tilde m_\Delta^\ell + \frac{\tilde m_\Delta^2}
    {4\,\tilde m_\Delta^\ell}\right)}\,,
\end{equation}
where the parameters $\tilde m_\Delta^\ell$ and $\tilde m^2_\Delta$ are given by
\begin{equation}
  \label{eq:definition-mtildes}
  \tilde m_\Delta^\ell=\frac{v^2\,|\pmb{Y}|^2}{M_\Delta}
  \qquad\mbox{and}\qquad
  \tilde m^2_\Delta=\mbox{Tr}[\pmb{m_\nu^{II}}\pmb{m_\nu^{II}}^\dagger]\,,
\end{equation}
with $|\pmb{Y}|^2=\mbox{Tr}[\pmb{Y}\,\pmb{Y}^\dagger]$.  In these
definitions we have replaced the trilinear coupling $\mu$ by the
contribution of the type-II sector to the effective light neutrino
mass matrix, encoded in $\tilde m^2_\Delta$. In principle this is just
a matter of choice, but it proves to be quite convenient given that in
contrast to $\mu$ the parameter $\tilde m_\Delta$ is (partially)
constrained by experimental neutrino data.

As we have already stressed the scenario we consider here is
defined by a hierarchical RH neutrino mass spectrum and a mild
hierarchy between $M_{N_1}$ and $M_\Delta$, that we choose to be in
the range $10^{-1}\lesssim M_\Delta/M_{N_1}\lesssim 1$. Consequently,
as explained at the beginning of this section $z=M_{\Delta}/T$ and
$z_N=r\,z$ with $r=M_{N_1}/M_\Delta$. In that way, once the CP
asymmetries in both sectors (I and II) are specified, the problem of
studying the evolution of the $L$ asymmetry is completely determined
by five parameters: the contribution of the lightest RH neutrino to
light neutrino masses $\tilde
m_{N_1}=v^2(\pmb{\lambda}\pmb{\lambda}^\dagger)_{11}/M_{N_1}$, $\tilde
m_\Delta$, $\tilde m_\Delta^\ell$, $M_\Delta$ and $r$ \footnote{This
  is to be compared with the pure triplet leptogenesis scenario
  \cite{Hambye:2005tk} where the generation of the $L$ asymmetry is
  entirely determined by only three parameters: $\tilde m_\Delta$,
  $\tilde m_\Delta^\ell$, $M_\Delta$.}.

Assuming an initial vanishing $L$ asymmetry ($Y_{\Delta_L}(z\to
0)=0$), the formal solution of the differential equation governing this
asymmetry in eq.  (\ref{eq:relevant-BEQs}) yields
\begin{equation}
  \label{eq:BmL-formal-sol}
  Y_{\Delta_L}(z)=\int_{z_i}^z\,dz'\,P(z')\,e^{-\int_{z'}^z\,dz''\,Q(z'')}\,,
\end{equation}
with the functions $P(z)=P^I(z) + P^{II}(z)$ and $Q(z)$ given by
\begin{align}
  \label{eq:PandQ-functions1}
  P^I(z)&=\frac{1}{s(z)H(z)z}
  \left\{
    \left[
      (y_{N_1}(z)-1)\epsilon_{N_1}^\text{tot}-y_{\Delta_\Delta}^H(z)
    \right]\gamma_{D_{N_1}}(z)
  \right\}\,,\\
  \label{eq:PandQ-functions2}
  P^{II}(z)&=\frac{3}{s(z)H(z)z}
  \left\{
    \left[
      (y_\Sigma(z) - 1)\epsilon_\Delta - 2 K_\ell\,y_{\Delta_\Delta}(z)
    \right]\gamma_{D_\Delta}(z)
  \right\}\,,\\
  \label{eq:PandQ-functions3}
  Q(z)&=\frac{1}{s(z)H(z)z}
  \left[
    \frac{1}{Y^\text{Eq}_\ell}
    \left(
      \gamma_{D_{N_1}}(z) + 2\,K_\ell\,\gamma_{D_\Delta}(z)
    \right)
  \right]\,.
\end{align}
Note that in $P^{II}(z)$ we have included a factor of 3 coming from
the triplet $SU(2)$ physical degress of freedom. By factorizing either
$\epsilon_{N_1}^\text{tot}$ or $\epsilon_\Delta$ from the functions
$P^{I,II}(z)$ and normalizing to $Y^\text{Eq}_\text{tot}\equiv
Y^\text{Eq}_\text{tot}(z\to0)=
Y^\text{Eq}_{N_1}(z)+Y^\text{Eq}_\Sigma(z)|_{z\to 0}$ the $L$
asymmetry in (\ref{eq:BmL-formal-sol}) can be written in terms of
efficiency functions that depend on the dynamics of the scalar triplet
and the fermionic singlet, namely
\begin{equation}
  \label{eq:efficiencies}
  Y_{\Delta_L}(z)=-\epsilon_{N_1}^\text{tot}\,Y^\text{Eq}_\text{tot}\,\eta^I(z)
  \quad\mbox{and}\quad
  Y_{\Delta_L}(z)=-\epsilon_\Delta\,Y^\text{Eq}_\text{tot}\,\eta^{II}(z)\,.
\end{equation}
The functions $\eta^{I,II}(z)$ are defined in such a way that in the
limit in which the triplet (RH neutrino) interactions are absent
$\eta^{I}$ ($\eta^{II}$) corresponds to the efficiency function of
standard leptogenesis (pure triplet leptogenesis). As usual the final
$L$ asymmetry is obtained from these functions in the limit $z\to
\infty$.

In models with an interplay of type I and II seesaws exhibiting a mild
hierarchy between the $\Delta$ and $N_1$ masses several scenarios for
the generation of the lepton asymmetry can be considered:
\begin{enumerate}[I.]
\item \label{item:lepto-case1}\underline{Purely triplet scalar leptogenesis models:}\\
  The relevant parameters follow the hierarchy $\tilde m_{N_1}\ll
  \tilde m_\Delta^\ell, \tilde m_\Delta$. The $L$ asymmetry is
  generated through the processes $\pmb{\Delta}\to \bar\ell\bar\ell$
  or $\pmb{\Delta}\to H H$ and the details strongly depend on whether
  $\tilde m_\Delta^\ell\gg \tilde m_\Delta$, $\tilde m_\Delta^\ell\ll
  \tilde m_\Delta$ or $\tilde m_\Delta^\ell\sim \tilde m_\Delta$.
  Interestingly, when $\tilde m_\Delta^\ell\gg \tilde m_\Delta$ the
  Higgs asymmetry -being weakly washed out- turns out to be large and
  implies a large lepton asymmetry.

  In order to illustrate how leptogenesis proceeds in these models we
  have solved the network of Boltzmann equations in
  (\ref{eq:relevant-BEQs}) for the benchmark point $P_I$=($\tilde
  m_{N_1}$, $\tilde m_\Delta$, $\tilde m_\Delta^\ell$,$M_\Delta$,$r$)
  =($10^{-4}$ eV, $10^{-2}$ eV, $10^{-1}$ eV, $10^{10}$ GeV,2) for
  fixed $\epsilon_\Delta=10^{-6}$ and $\epsilon_{N_1}=10^{-5}$ and
  assuming initial vanishing asymmetries. The results are displayed in
  fig. \ref{fig:models-case1}. With this parameter choice the
  generation of the $L$ is entirely determined by the triplet
  reactions. As can be seen in fig. \ref{fig:models-case1}
  (left-panel) the RH neutrino reaction density is tiny implying in
  the equation for the evolution of the $L$ asymmetry in
  (\ref{eq:relevant-BEQs}) the RH related quantities can be neglected.
  For the triplet-Higgs reaction density $\gamma_{D_\Delta}^H$, though
  also small, this is not the case: the relevant reaction density in
  the pure type II sector is
  $\gamma_{D_\Delta}=\gamma_{D_\Delta}^H+\gamma_{D_\Delta}^\ell$ so
  due to the constraint $\gamma_{D_\Delta}^H\ll\gamma_{D_\Delta}^\ell$
  the Higgs asymmetry does not suffers from a strong washout ($K_H\sim
  10^{-3}$) and becomes large.  Due to the hypercharge neutrality
  condition, eq. (\ref{eq:hypercharge-neutrality}), a large lepton and
  triplet asymmetry develops. Once the triplet number density starts
  being diluted, due to decays controlled by $\gamma_{D_\Delta}^\ell$,
  the asymmetry stored in triplets is transferred to the lepton
  asymmetry. This can be observed in fig. \ref{fig:models-case1}
  (right-panel) where at $z\sim 5$ the triplet asymmetry drops and the
  $Y_{\Delta_L}$ increases accordingly. The situation described here
  is quite similar to what happens in standard flavored leptogenesis
  \cite{hep-ph/0601084,hep-ph/0605281,hep-ph/9911315} where, depending
  on the lepton flavor projectors, a large asymmetry in a particular
  flavor can be stored yielding in turn a large net lepton asymmetry.
\begin{figure}
  \centering
  \includegraphics[width=7.4cm,height=6cm]{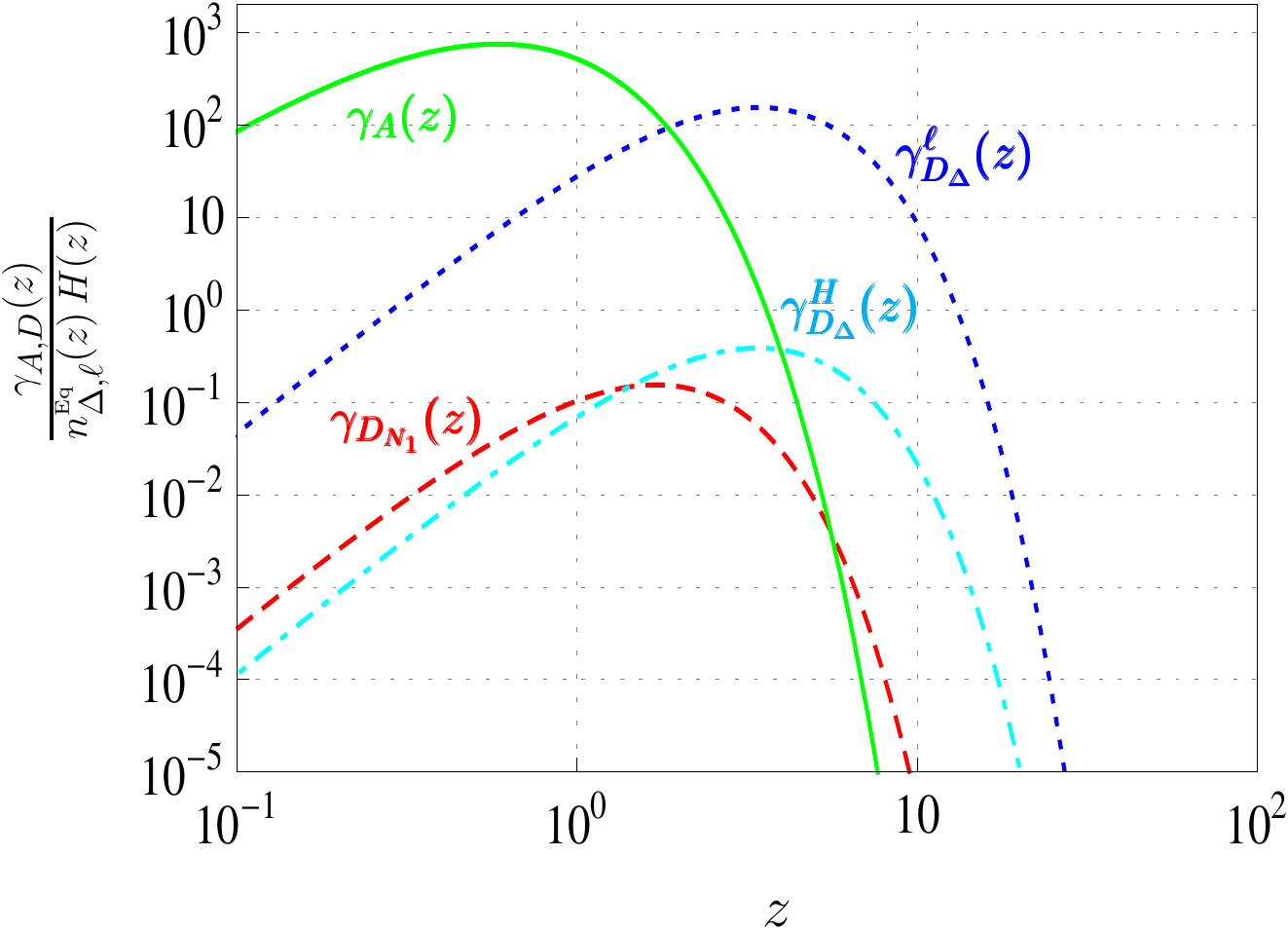}
  \includegraphics[width=7.4cm,height=6cm]{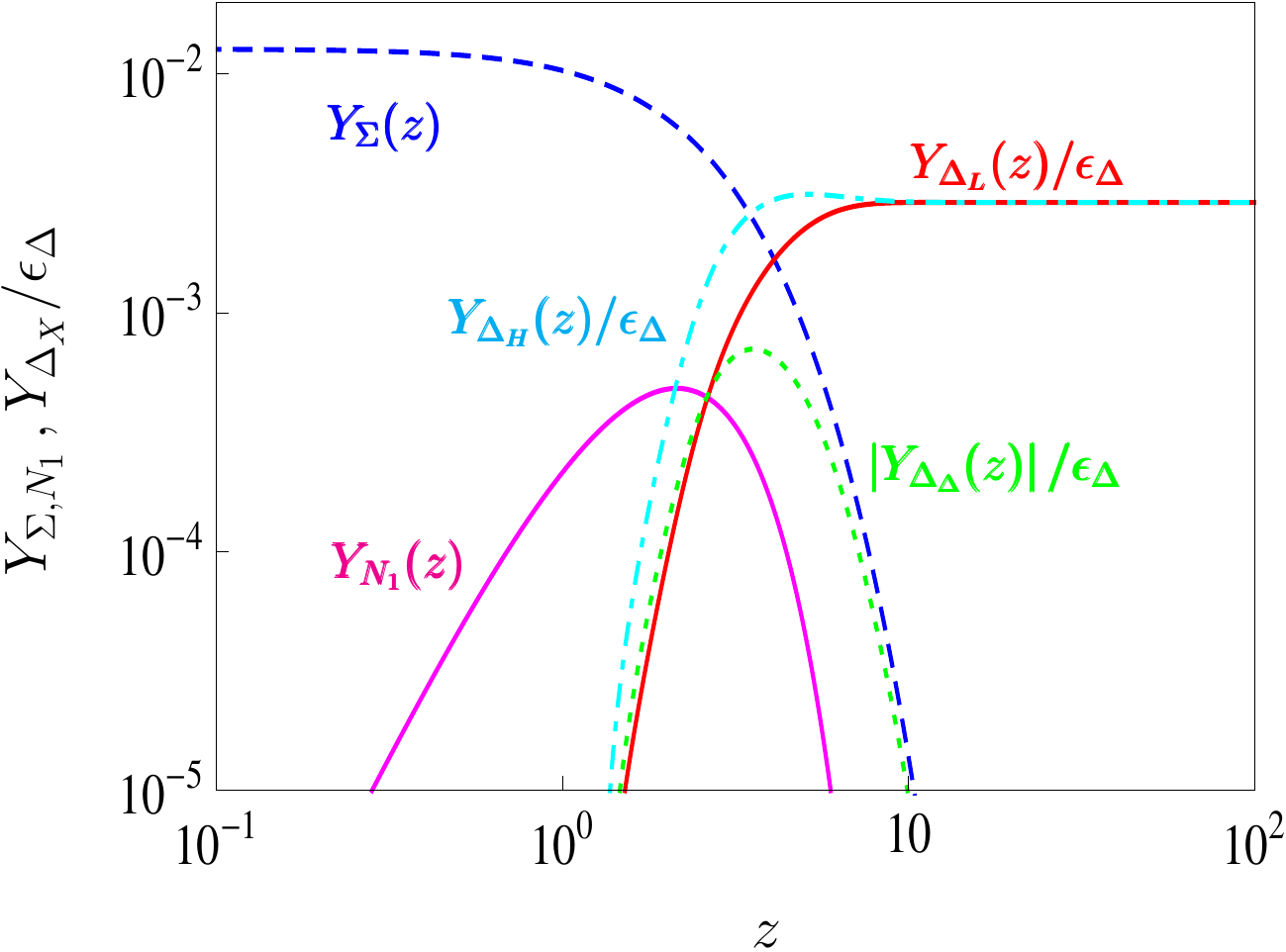}
  \caption{{\it Reaction densities for triplet and RH neutrino
      processes (left-panel) and evolution of the different densities
      (right-panel) entering in the kinetic equations for the scenario
      of purely triplet leptogenesis. See the text for more details.}}
  \label{fig:models-case1}
\end{figure}
\item\label{item:lepto-case2} \underline{Singlet dominated leptogenesis models:}\\
  These scenarios are defined according to $\tilde m_{N_1}\gg \tilde
  m_\Delta^\ell, \tilde m_\Delta$ thus leptogenesis is mainly
  determined by $N_1$ dynamics. The relative difference between the
  parameters $\tilde m_\Delta^\ell$ and $\tilde m_\Delta$ determines
  whether either the Higgs asymmetry or the $L$ asymmetry are strongly
  or weakly washed out, thus three cases can be distinguished: $\tilde
  m_\Delta^\ell\gg \tilde m_\Delta$, $\tilde m_\Delta^\ell\ll \tilde
  m_\Delta$ or $\tilde m_\Delta^\ell\sim \tilde m_\Delta$. Each of
  them exhibit different features.

  \begin{figure}
    \centering
    \includegraphics[width=7.4cm,height=6cm]{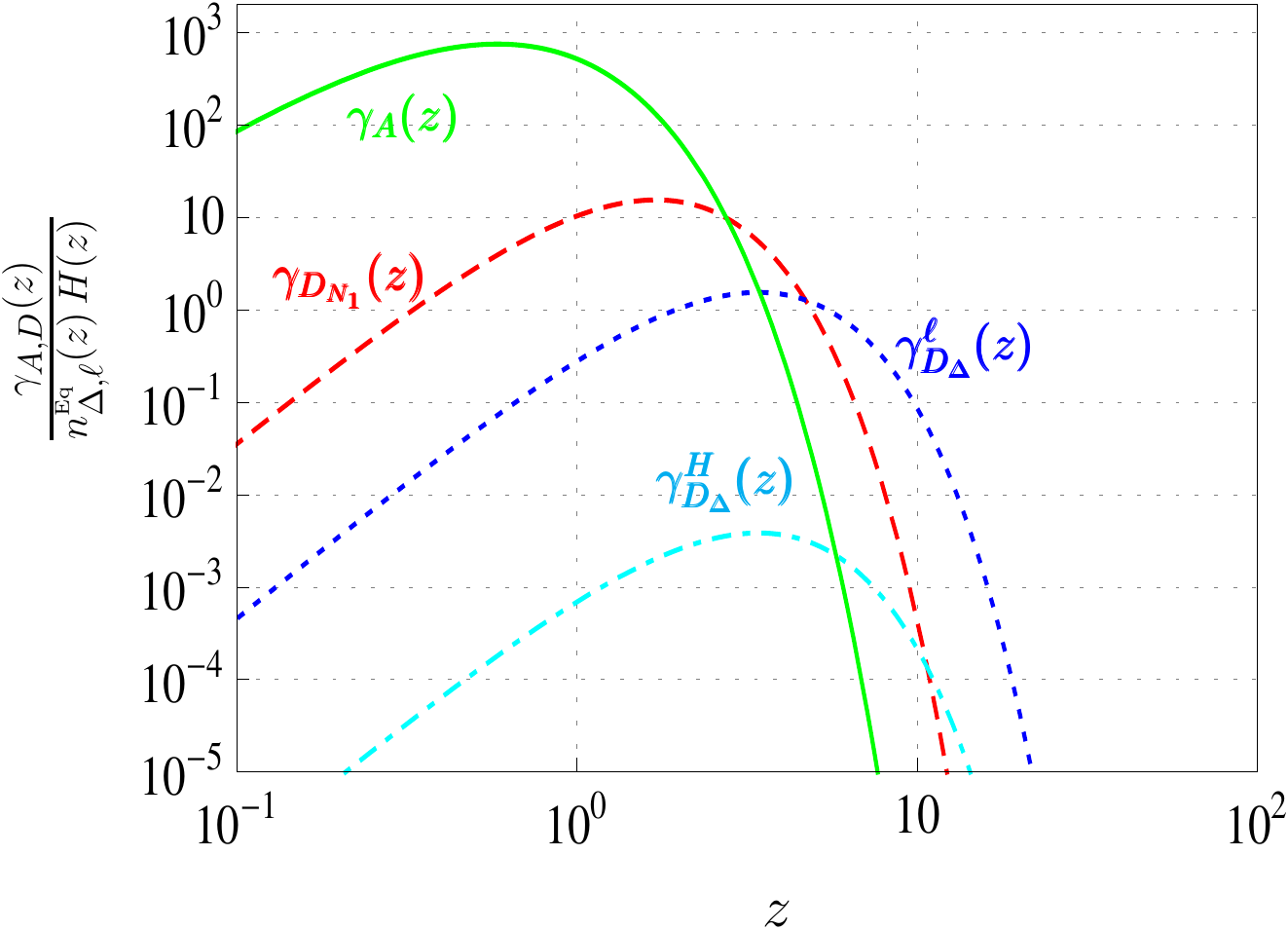}
    \includegraphics[width=7.4cm,height=6cm]{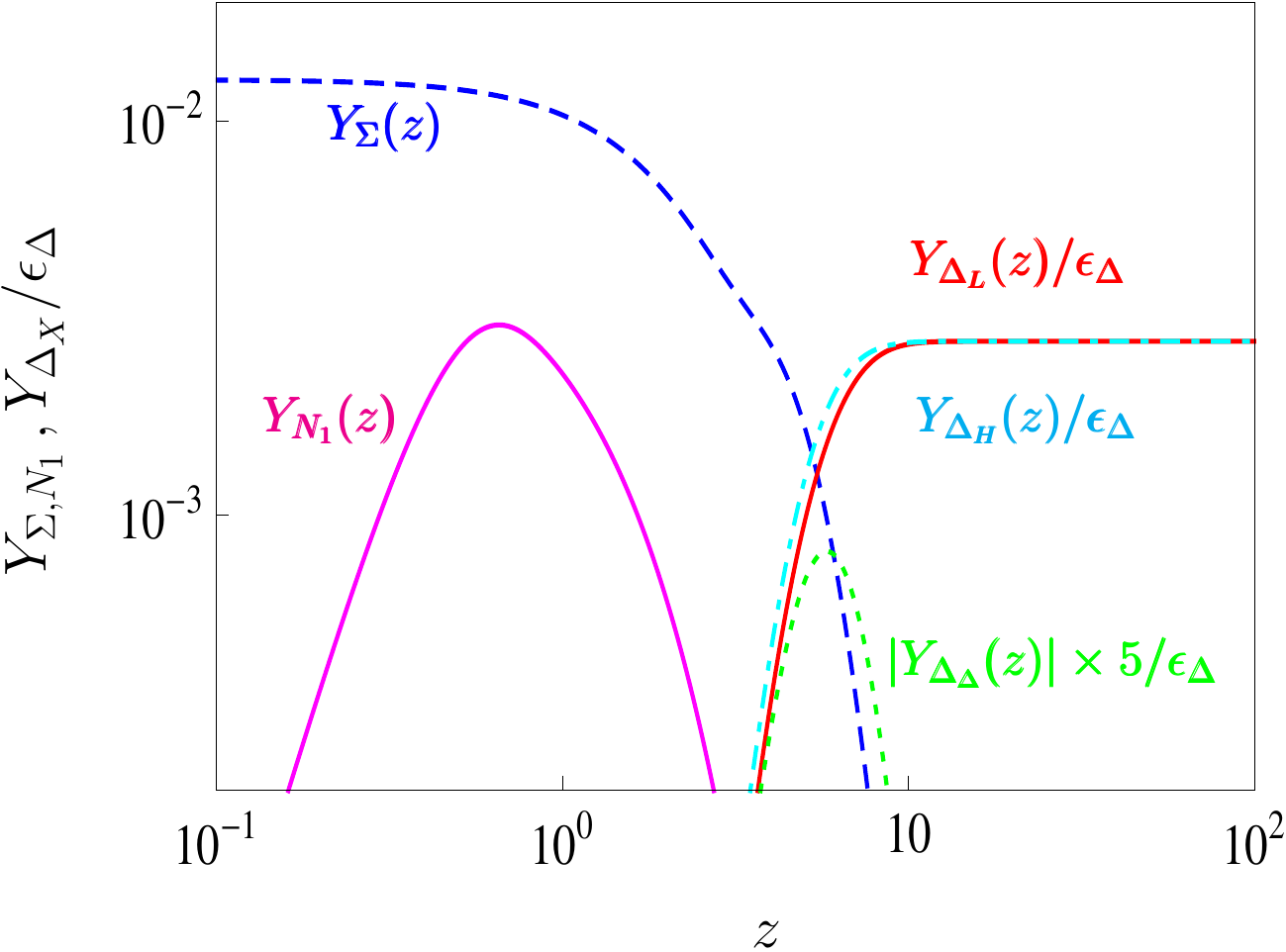}
    \caption{{\it Reaction densities for triplet and RH neutrino
        processes (left-panel) and evolution of the different densities
        (right-panel) entering in the kinetic equations for the scenario
        of singlet dominated leptogenesis. See the text for more details.}}
    \label{fig:models-case2}
  \end{figure}
  We have solved eqs.~(\ref{eq:relevant-BEQs}) for the case $\tilde
  m_\Delta^\ell\gg \tilde m_\Delta$ by fixing the parameter space
  point $P_{II}$=($\tilde m_{N_1}$, $\tilde m_\Delta$, $\tilde
  m_\Delta^\ell$,$M_\Delta$,$r$) =($10^{-2}$ eV, $10^{-4}$ eV,
  $10^{-3}$ eV, $10^{10}$ GeV,2) with the CP asymmetries
  $\epsilon_{N_1}^\text{tot},\epsilon_\Delta$ settled as in the
  previous analysis. The results are displayed in figure
  \ref{fig:models-case2} (right-panel). Due to the hierarchy
  $\gamma_{D_{N_1}}\gg \gamma_{D_\Delta}^{\ell, H}$ (see
  fig. \ref{fig:models-case2} (left-panel)) the triplet asymmetry is
  tiny. The hypercharge neutrality condition thus implies almost
  exact $L$ and Higgs asymmetries, the deviations determined by the
  small amount of $Y_{\Delta_\Delta}$ produced. Compared with the
  models discussed in the previous item in this case $Y_\Sigma$
  departs from thermal equilibrium at higher temperatures (smaller
  values of $z$), the decoupling temperature entirely determined by gauge
  reaction decoupling ($\gamma_A(z)/(n_\Delta^\text{Eq}(z)H(z))<1$) \footnote{Note
    that in contrast to the previous case the Higgs and Yukawa triplet
    related reactions are always decoupled.}. For the example in
  figure \ref{fig:models-case2} it can be seen this happens at $z\sim
  5$, the small bump in $Y_\Sigma(z)$ is actually due to this effect.
\item\label{item:lepto-case3}\underline{Mixed leptogenesis models:}\\
  In these models the parameters controlling gauge reactions strengths
  are all of the same order i.e.  $\tilde m_{N_1}\sim \tilde
  m_\Delta^\ell\sim \tilde m_\Delta$, so drawing general conclusions
  from the analysis of particular benchmark points is far more
  involved and the general picture may only be obtained by scans of
  the relevant parameters. However, with the purpose of highlighting
  some of the aspects of these schemes we have solved the set of
  eqs. in (\ref{eq:relevant-BEQs}) assuming $P_{III}$=($\tilde
  m_{N_1}$, $\tilde m_\Delta$, $\tilde m_\Delta^\ell$,$M_\Delta$,$r$)
  =($2\times 10^{-2}$ eV, $4\times 10^{-2}$ eV, $6\times 10^{-2}$ eV,
  $10^{10}$ GeV,2) with $\epsilon_{N_1}^\text{tot}$ and
  $\epsilon_\Delta$ fixed as in the previous two analyses. The results
  are shown in fig. \ref{fig:models-case3}, we have not included in
  this case neither the $\Sigma$ nor the $N_1$ abundances as the
  former behaves as in purely triplet leptogenesis models while the
  later as in singlet dominated ones. It can be seen that due to the
  parameter choice $\gamma_{D_\Delta}^{\ell,H}$ overcome the
  corresponding gauge reactions at $z\sim 3$ maintaining the
  $Y_\Sigma$ abundance close to the equilibrium distribution.  The
  total $L$ asymmetry is two orders of magnitude smaller than in the
  purely triplet and singlet dominated leptogenesis models we have
  considered, in part because the washouts in both the Higgs and
  lepton directions are large ($K_\ell=0.64 $, $K_H=0.36$) but it
  remains to be seen whether this is a general feature of these type
  of models.
  \begin{figure}
    \centering
    \includegraphics[width=7.4cm,height=6cm]{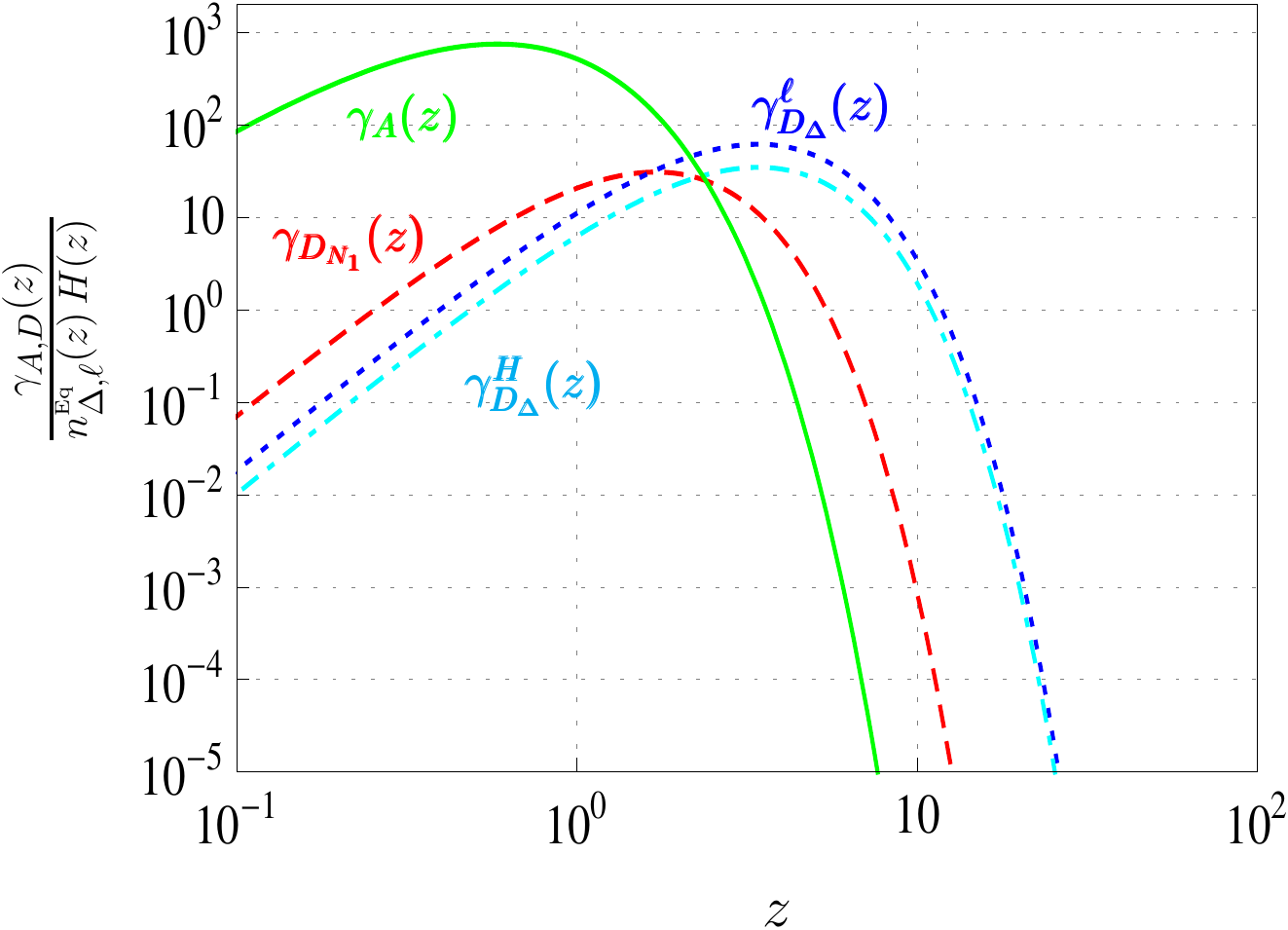}
    \includegraphics[width=7.4cm,height=6cm]{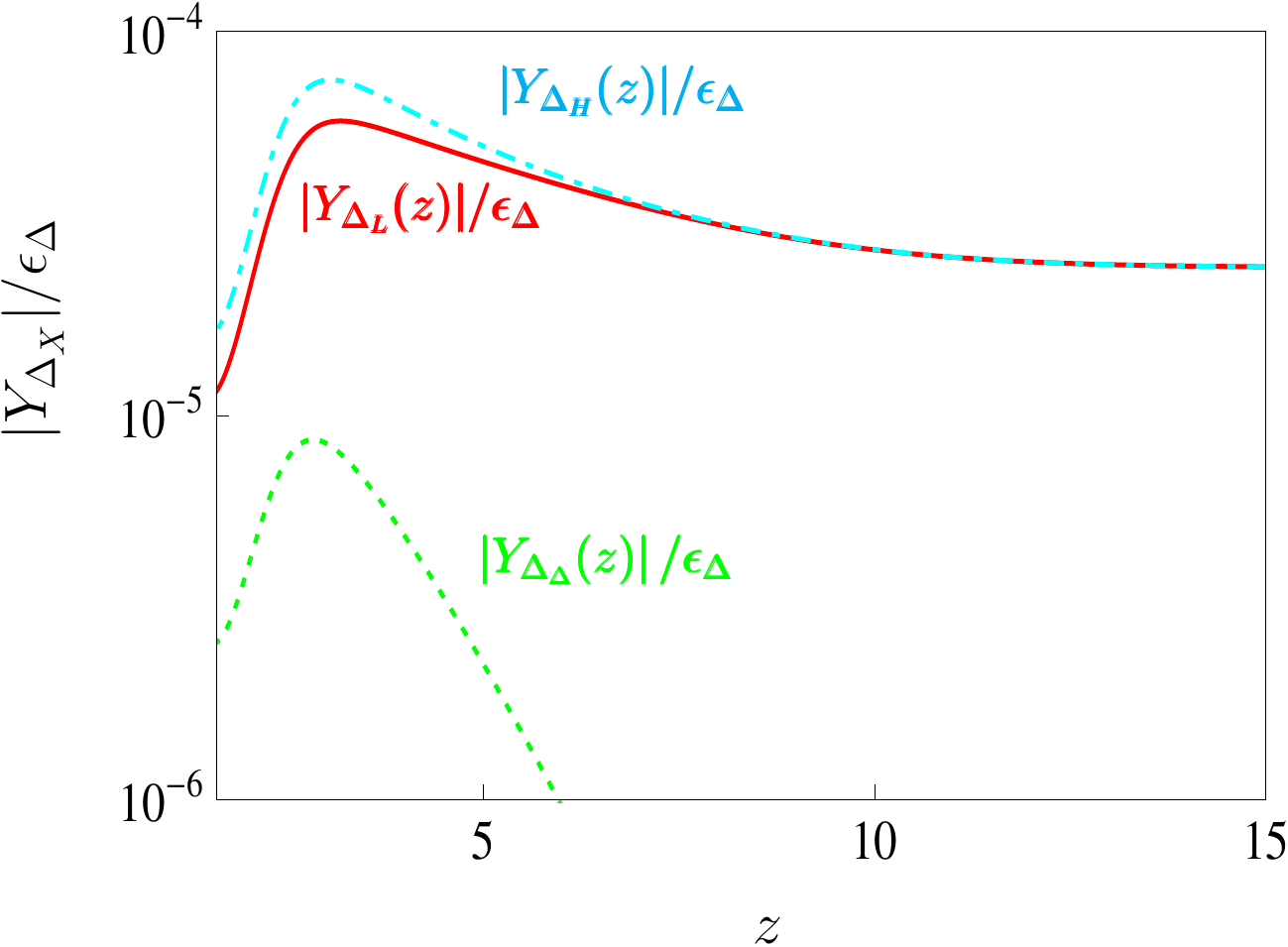}
    \caption{{\it Reaction densities for triplet and RH neutrino
        processes (left-panel) and evolution of the different
        densities (right-panel) entering in the kinetic equations for
        mixed leptogenesis models. See the text for more details.}}
    \label{fig:models-case3}
  \end{figure}
\end{enumerate}

\section{Conclusions}
\label{sec:conclusions}
We have studied leptogenesis in models featuring the interplay between
type I and type II seesaw in the presence of flavor symmetries. By
assuming an exact mixing scheme we classified in a model independent
way the possible realizations of such hybrid scenarios. The possible
models can be grouped according to whether in the limit of exact
mixing the CP violating asymmetry $\epsilon_N$ vanishes or not. Our
results show that the presence of additional degrees of freedom may
allow viable leptogenesis without requiring any departure from exact
mixing, particularly for non-minimal models that can include the
identity matrix (according to the definition of minimal in
sec. \ref{sec:exact}). Effectively, only the repetition of eigenvectors across both seesaws can produce a contribution and if no eigenvectors are repeated the asymmetry must vanish - however the identity matrix counts as all eigenvectors so its presence on one of the seesaws automatically repeats any present in the other. In the case of minimal models the constraints
imposed by exact mixing enforce a vanishing CP violating asymmetry,
thus resembling the results of flavor models based just on type I
seesaw.

As regards the generation of the lepton asymmetry we have shown that
by going beyond the limit of RH neutrino decoupling, allowing the RH
neutrino and triplet mass splitting to be small, permits a
variety of scenarios for leptogenesis. We have done an analysis of
these realizations for particular benchmark points in the relevant
parameter space.  Our results show these models exhibit different
qualitative as well as quantitative features that may lead
leptogenesis to be quite different from the standard picture.

Although based on the TB mixing scheme -for concreteness- our
conclusions remain valid regardless of the mixing pattern, provided
the flavor symmetry enforces the type I and type II light neutrino
mass matrices to be form-diagonalizable. Accordingly, our findings
constitute a new pathway to leptogenesis in models in which a flavor
symmetry is responsible for lepton mixing.
\section{Acknowledgments}
We would like to thank Stefan Antusch for helpful comments.
DAS is supported by a Belgian FNRS postdoctoral fellowship.
IdMV is supported by DFG grant PA 803/6-1.
\appendix
\section{Notation and conventions}
\label{sec:appendix}
In this appendix we fix the conventions that have been used in 
sec.~\ref{sec:BmLasymmetry} for the discussion of the different
leptogenesis realizations one can define in models featuring an
interplay between type I and II seesaws and a mild hierarchy between
the electroweak triplet scalar and the lightest RH neutrino.

We have used Maxwell-Boltzmann distribution functions for massless
(relativistic) as well as massive species:
\begin{equation}
  \label{eq:number-dens-distributions}
  n_\Delta^\text{Eq}(z)=\frac{3\,M_\Delta^3}{2\,\pi^2}\,\frac{K_2(z)}{z}\,,
  \quad
  n_{N_1}^\text{Eq}(z)=\frac{M_\Delta^3}{\pi^2}\,r^2\frac{K_2(r\,z)}{z}\,,
  \quad
  n_{\ell,H}^\text{Eq}(z)=\frac{2}{\pi^2}\,\frac{M_\Delta^3}{z}\,,
\end{equation}
where, as mentioned in sec. \ref{sec:BmLasymmetry},
$r=M_{N_1}/M_\Delta$ and $K_n(z)$ is the $n$-th order modified Bessel
function of the second type. The entropy density and the expansion 
rate of the Universe are defined as usual, namely
\begin{equation}
  \label{eq:entropy-dens}
  s(z)=\frac{4\,g_*}{\pi^2}\frac{M_\Delta^3}{z^3}\,,\quad
  H(z)=\sqrt{\frac{8 g_*}{\pi}}\frac{M_\Delta^2}{M_\text{Planck}}\frac{1}{z^2}
\end{equation}
with the number of relativistic degrees of freedom given by $g_*=118$
and the Planck mass by $M_\text{Planck}=1.22\times 10^{19}$ GeV. 

The reaction densities for $1\leftrightarrow 2$ processes are given by
\begin{align}
  \label{eq:reac-dens-1to2}
  \gamma_{D_{N_1}}(z)&=\frac{1}{8\,\pi^3}\frac{M_\Delta^5}{v^2}r^4
  \frac{K_1(rz)}{z}\,\tilde m_{N_1}\,,
  \nonumber\\
  \gamma_{D_\Delta}^H(z)&=\frac{1}{8\,\pi^3}\frac{M_\Delta^5}{v^2}
  \frac{K_1(z)}{z}\,\frac{\tilde m_\Delta^2}{4\tilde m_\Delta^\ell}\,,
  \nonumber\\
  \gamma_{D_\Delta}^\ell(z)&=\frac{1}{8\,\pi^3}\frac{M_\Delta^5}{v^2}
  \frac{K_1(z)}{z}\,\tilde m_\Delta^\ell\,,
\end{align}
where the full triplet reaction density is given by
$\gamma_{D_\Delta}=\gamma_{D_\Delta}^H+\gamma_{D_\Delta}^\ell$.  For
$2\leftrightarrow 2$ triplet annihilations the corresponding reaction density
reads
\begin{equation}
  \label{eq:reac-dens-2to2}
  \gamma_A(z)=\frac{M_\Delta^4}{64\,\pi^4}\,\int_4^\infty
  dx\sqrt{x}\frac{K_1(zx)}{z}\,\widehat\sigma_A(x)\,,
\end{equation}
where $x=s/M_\Delta^2$. The reduced cross section $\widehat\sigma_A(x)=2\,x\,\lambda(1,x^{-1},0)$
(where we have $\lambda(a,b,c)=(a-b-c)^2-4bc$) and involves the $s$-channel processes
$\pmb{\Delta}\pmb{\Delta}\to F\bar F, AA, HH$ ($F$ and $A$ stand for
standard model fermions and $SU(2)\times U(1)$ gauge bosons
respectively), $t$ and $u$ channel triplet mediated processes
$\pmb{\Delta}\pmb{\Delta}\to AA$ and the ``quartic'' process
$\pmb{\Delta}\pmb{\Delta}\to AA$. In powers of the kinematic factor
$\omega(x)=\sqrt{1-4/x}$ it can be split in three pieces
\cite{Hambye:2005tk}:
\begin{align}
  \label{eq:reduced-cross-section}
  \widehat \sigma_1(x)&=\frac{1}{\pi}
  \left[
    g^4\left(5 + \frac{34}{x}\right)
    +
    \frac{3}{2}g'^4\left(1 + \frac{4}{x}\right)
  \right]\omega(x)\,,
  \nonumber\\
  \widehat \sigma_2(x)&=\frac{1}{8\pi}
  \left(
    25 g^4 + \frac{41}{2}g'^4
  \right)\omega(x)^3\,,
  \nonumber\\
  \widehat \sigma_3(x)&=\frac{6}{\pi x^2}
  \left[
    4 g^4 (x-1) + g'^4 (x-2)
  \right]\ln\left[
    \frac{1+\omega(x)}{1-\omega(x)}
  \right]
  \,,
\end{align}
with $\widehat\sigma_A(x)=\sum_{i=1}^3\widehat \sigma_i(x)$ and $g,
g'$ the $SU(2)$ and $U(1)$ gauge couplings.

\end{document}